\documentclass[twocolumn,showpacs,preprintnumbers,amsmath,amssymb]{revtex4}

\usepackage{epsfig}
\usepackage{psfig}

\newcommand{\vektor}{\boldsymbol}
\newcommand{\hatR}{\hat{\mathrm R}}
\newcommand{\gammafktB}[2]
{\frac{\Gamma\big(\frac{#1}{2}\big)}{\Gamma\big(\frac{#2}{2}\big)}}
\newcommand{\JacP}{{\mathcal P}}
\newcommand{\gammafkt}[1]
{\Gamma\big(\frac{#1}{2}\big)}

\begin{document}


\title{Semi-analytic Faddeev solution to the $N$-boson problem with 
zero-range interactions}

\author{T. Sogo}
\author{O. S{\o}rensen}
\author{A.S. Jensen}
\author{D.V. Fedorov}
\affiliation{ Department of Physics and Astronomy,
        University of Aarhus, DK-8000 Aarhus C, Denmark }

\date{\today}

\begin{abstract}
We study two-body correlations for $N$ identical bosons by use of the
hyperspherical adiabatic expansion method.  We use the zero-range
interaction and derive a transcendental equation determining the key
ingredient of the hyperradial potential. The necessary renormalization is
for both repulsive and attractive interactions achieved with an
effective range expansion of the two-body phase-shifts.  Our solutions
including correlations provide the properties
of Bose-Einstein condensates exemplified by stability conditions as
established by mean-field Gross-Pitaevskii calculations.
The many-body Efimov states are unavoidable for large scattering
lengths.
\end{abstract}


\maketitle

\paragraph*{Introduction.}

Solutions to the two-body problem are described in many textbooks.
The three-body problem is solvable in practice at least for
short-range interactions \cite{nie01,mac02}.  Analytic solutions are found
for specific potentials, e.g. square-well potentials confined to only
$s$-waves, where the large distance behavior is particularly simple
\cite{jen97}.  This limit is effectively the result for a zero-range
interaction which is accurate when the small distances comparable
to the interaction range are uninteresting or only needed to provide
overall average properties.  Therefore much effort has been devoted to
the three-body problem by direct use of zero-range interactions, where
regularization is needed to prevent unphysical collapsed wave
functions \cite{bed99,fed01,yam04}.

Not surprising the $N$-body problem has not been solved in general.
The most popular approximation is the mean-field assumption which
minimizes the technical difficulties but ignores correlations.  Even
then often the zero-range interaction is employed to exploit the
additional simplifications allowing elaborate systematic investigations.
Two prominent examples could be the Skyrme-Hartree-Fock fermion
calculations for $N$-nucleon systems, see e.g.\cite{dob04}, and the
analogous Gross-Pitaevskii calculations for dilute atomic Bose gases, see
e.g. \cite{esr99,abd01}.

Improvements within the mean-field approximation require better
interactions, for example of finite range. This changes significantly
solutions of high density as for nuclei but would not be visible for
dilute condensed atomic gases.  Another improvement to include
correlations and go beyond the mean-field is usually much more
difficult. For nuclei a perturbative treatment starting with the
zero-range interaction and the mean-field solution is at the moment
not very useful since a collapse into cluster components only can be
avoided by a large additional phenomenological renormalization of the
interaction.  For dilute systems like atomic gases an appropriately
renormalized zero-range interaction is not a severe approximation
\cite{ols02} and it would therefore be directly useful to include
effects of correlations.

Transparent or even analytic extension beyond the mean-field
approximation is highly desirable for several reasons, e.g. (i)
provide simplicity and insight, (ii) allow systematics and access of
complicated systems, (iii) provide intermediate and large distance
properties, which in quantum mechanics often are directly responsible
for the qualitative features of the entire solution, (iv) in designs
of more efficient numerical methods by use of the available large
distance asymptotic.  The purpose of the present letter is to lay out
the foundation for a series of applications and extensions by
providing a semi-analytic solution to the $N$-boson problem.

\paragraph*{Theory.}

We consider a dilute system of $N$ identical bosons of mass $m$, coordinates
$\vektor r_{i}$ and momenta $\hat {\vektor p}_i$.  They affect each
other via a two-body short-range interaction $V(r)$ which can be
approximated by a zero-range potential. The Hamiltonian is then 
\begin{eqnarray} \label{e30}
  \hat H  =
  \sum_{i=1}^N
  \bigg(
  \frac{\hat {\vektor p}_i^2}{2m}+\frac12m\omega^2r_i^2
  \bigg)
  +\sum_{i<j}^N V(\vektor r_{ij})
  \;,
\end{eqnarray}
where $r_{ij}=|\vektor r_j-\vektor r_i|$ and an external harmonic
field of angular frequency $\omega$ is added.  We choose the
hyperspherical adiabatic expansion method where the principal
coordinate is the hyperradius $\rho$ defined by \cite{nie01,sor02a}
\begin{eqnarray}  \label{e40}
 \rho^2=\frac{1}{N}\sum_{i<j}^N r_{ij}^2 = 
 \sum_{i}^N (\vektor r_{i} - \vektor R)^2
 =  \sum_{i}^N r_{i}^2  -  N R^2  \;,
\end{eqnarray}
where $\vektor R$ is the center of mass.  The remaining degrees of
freedom are the dimensionless hyperangles, collectively denoted by
$\Omega$.  The lowest adiabatic relative wave function $\Psi$ is given
by
\begin{eqnarray}  \label{e50}
  \Psi(\rho,\Omega)  =  
  \frac{f(\rho) \Phi(\rho,\Omega) }{\rho^{(3N-4)/2}}   \;\;,\;\;
  \Phi(\rho,\Omega)=  \sum_{i<j}^N \phi(\rho,r_{ij})  \;,
\end{eqnarray}
where the angular part $\Phi$ is expressed as a sum of two-body
correlation amplitudes, Faddeev components, 
$\phi$ each assumed to depend only on the overall
size $\rho$ and the distance between the pairs of particles. For
zero-range interactions this dependence is no further restriction as
two particles then only interact via $s$-waves.  For fixed hyperradius
the free variable in $r_{ij}$ can conveniently be substituted by an
angle $\alpha_{ij}$ defined by $r=\sqrt2\rho\sin\alpha$, where we
omitted the indices $ij$.

The Faddeev component $\phi(\alpha) \equiv
\phi(\rho,\sqrt2\rho\sin\alpha)$ is determined from the angular
Faddeev equation \cite{sor02a,fad61}, i.e.
\begin{eqnarray}  \label{e60}
  0
  =
  \big[
  \hat\Pi^2+v(\alpha)\hatR -  \lambda(\rho)
  \big]
  \phi(\alpha)
  \;,
\end{eqnarray}
where $\hbar^2\lambda/(2m\rho^2)$ is the energy eigenvalue,
$v(\alpha)$ is related to the two-body potential $V(r)$ by $v(\alpha)
= 2m\rho^2V(\sqrt2\rho\sin\alpha)/\hbar^2$, and the kinetic-energy,
$\hat\Pi^2$, and rotation, $\hatR$, operators are given by
\begin{eqnarray}  \label{e70}
  &&
  \hat\Pi^2  = 
  -\frac{\partial^2}{\partial\alpha^2}+
 \frac{3N-9-(3N-5)\cos 2\alpha}{\sin 2 \alpha}
  \frac{\partial}{\partial\alpha}
  \;, \\  \label{e80}
  &&
  \hatR
  =
  1
  +2(N-2)\hatR_{13}
  +\frac12(N-2)(N-3)\hatR_{34}
  \;.
\end{eqnarray}
The operators, $\hatR_{13}$ and $\hatR_{34}$, in eq.~(\ref{e80})
``rotate'' two-body Faddeev components 
between particles $1$ (or $2$) and a third particle ($2(N-2)$ terms), 
and ones between particles $3$ and $4$ both different from $1$ and $2$ 
($(N-2)(N-3)/2$ terms).
The resulting radial equation is then
\begin{eqnarray}  \label{e110}
  \bigg[
  -\frac{\hbar^2}{2m}
  \frac{d^2}{d\rho^2}
  +U(\rho)-E\bigg]
  f(\rho)=0
  \;,
\end{eqnarray}
where $E$ is the energy and the effective radial potential is
\begin{eqnarray}  \label{e120}
  \frac{2mU(\rho)}{\hbar^2}
  \equiv
  \frac{\lambda(\rho)}{\rho^2}+
  \frac{(3N-4)(3N-6)}{4\rho^2}+
  \frac{\rho^2}{b_t^4}
  \;
\end{eqnarray}
with the oscillator length $b_t^2 \equiv \hbar/(m \omega)$.  This
potential is therefore determined by the external trap ($\rho^2$), the
generalized centrifugal barrier ($\rho^{-2}$) and the interaction part
($\lambda(\rho)$) from the angular equation.  The properties of the
eigenvalue $\lambda$ from eq.~(\ref{e60}) are then decisive for the
radial potential and the corresponding energies and wave functions.

\paragraph*{Zero-range interaction.}

We approximate $v$ by a zero-range interaction which means that
eq.~(\ref{e60}) has only  kinetic energy terms for all $\alpha \neq
0$. The solution is well known as the Jacobi functions
$\JacP_\nu^{(a,b)}(x)$, i.e.
\begin{eqnarray}  \label{e130}
  \phi_\nu(\alpha) &=&   \JacP_\nu^{(3N/2-4,1/2)}(-\cos2\alpha) \;,\;
 \\ \label{e135}  \lambda &=& 2\nu(2\nu+3N-5)  \;,
\end{eqnarray}
where the boundary condition $\phi_\nu(\alpha=\pi/2) = 0$ is fulfilled
by this Jacobi function \cite{nie01,sor02a}.  The boundary
condition of $\phi_\nu(\alpha=0)=0$ is only obeyed for these functions for
integer values of $\nu$ which therefore fully determines the free
solutions. However, at the point $\alpha=0$ we now have an infinitely
large potential which can be replaced by an appropriate boundary
condition, i.e. obtained by using the observation that the wave
function $\hatR \phi_\nu(\alpha)$ at small two-particle separation $r$
approaches the two-body wave function $u(r)$.  Then $\nu$ does not
have to be integer. The coordinate and wave function connections are
$r=\sqrt2\rho\sin\alpha \approx \sqrt{2}
\rho \alpha$ and $u(r) \propto \alpha \hatR \phi_\nu(\alpha)$. The boundary 
condition replacing the zero-range interaction for $u$ is \cite{fed01}
\begin{eqnarray} \label{e140}
  \frac{1}{u(r)}\frac{du(r)}{dr}\bigg|_{r=0}
  =
  -\frac{1}{a_s}
  +\frac12k^2R_{eff}
  +\mathcal{O}(k^4)  \;,
\end{eqnarray}
where $k$ is the wave number, $a_s$ is the scattering
length and $R_{eff}$ is the effective range.  For $R_{eff}=0$ we then
get
\begin{eqnarray}  \label{e150}
  \frac{\partial[\alpha \hatR \phi_\nu(\alpha)]}
  {\partial\alpha}\bigg|_{\alpha=0}  =
  -\frac{\sqrt{2}\rho}{a_s}
  \alpha \hatR \phi_\nu(\alpha)\bigg|_{\alpha=0}  \;.
\end{eqnarray}
For small $\alpha$ the solutions (\ref{e130}) behave as \cite{nie01}
\begin{eqnarray}   \label{e160}
  &&
  \phi_\nu(\alpha)
  \simeq
  \frac{A}{\alpha}+B
  \;,\quad
  A
  \equiv
  -\frac{\sin(\pi\nu)}{\sqrt\pi}
  \frac{\Gamma\big(\nu+\frac{3N-6}{2}\big)}{\Gamma\big(\nu+\frac{3N-5}2\big)}
  \;,\qquad\\   \label{e170}
  &&
  B
  \equiv
  \cos(\pi\nu)\frac{2}{\sqrt\pi}\frac{\Gamma\big(\nu+\frac32\big)}
  {\Gamma(\nu+1)}
  \;.
\end{eqnarray}
Then at the edge of the zero-range potential we get
\begin{eqnarray}   \label{e180}
  &&
  \alpha\hatR \phi_\nu(\alpha)\Big|_{\alpha=0}=A
  \;,\\    \label{e190}
  &&
  \frac{\partial[\alpha\hatR \phi_\nu(\alpha)]}
  {\partial\alpha}\bigg|_{\alpha=0}
  =B
  +(\hatR-1)\phi_\nu(0)
  \;.
\end{eqnarray}
Combining eqs.~(\ref{e150}), (\ref{e180}) and (\ref{e190}) we obtain 
\begin{eqnarray}   \label{e200}
  \frac{\rho}{a_s}
  =
  \frac{-1}{\sqrt2 A}
  \big[B+(\hatR-1)\phi_\nu(0)\big]
  \;.
\end{eqnarray}  
where $(\hatR -1)\phi_\nu(0)$ are given by eq.~(\ref{e80}) and the
explicit expressions
\begin{eqnarray}  \label{e90}
  &&
  \hatR_{34}\phi_\nu(0)  =
  \frac{4}{\sqrt\pi}\gammafktB{3N-6}{3N-9}  \Big(\frac{1}{2}\Big)^{(3N-6)/2}
 \\ \nonumber
 && \int_{-1}^{1}dx\; (1+x)^{1/2}  (1-x)^{(3N-11)/2}
  \JacP_\nu^{(3N/2-4,1/2)}(x) \;,\;
 \\   \label{e100}
  &&
  \hatR_{13}\phi_\nu(0) =
  \frac{2}{\sqrt\pi}\gammafktB{3N-6}{3N-9} \Big(\frac{2}{3}\Big)^{(3N-8)/2}
 \\ \nonumber
  && \int_{-1}^{1/2}dx\; (1+x)^{1/2}  (1/2-x)^{(3N-11)/2}
   \JacP_\nu^{(3N/2-4,1/2)}(x) \;.\;  
\end{eqnarray}  

The structure of eq.(\ref{e200}) is such
that the right hand side depends on the particle number $N$ and the
index $\nu$ on the Jacobi functions which in turn determines the
angular eigenvalue $\lambda$ from eq.(\ref{e130}).  Thus the left hand
side, $\rho/a_s$, is a unique function of $\nu$ and in turn of
$\lambda$ through eq.(\ref{e135}).  By inversion the effective radial
potential is determined as a function of hyperradius divided by the
scattering length.

Extension to finite values of $R_{eff}$ amounts to replacing $1/a_s$
in eq.(\ref{e200}) by $1/a_s - \frac12 k^2 R_{eff}$, where $m
E_2/\hbar^2 = k^2 = (\lambda + (9N-19)/2)/(2\rho^2)$ with the two-body
energy $E_2$. This substitution is obtained by eq.(\ref{e60}) in the
limit of small $\alpha$ from the connection between $u(r)$ and
$\phi_\nu(\alpha)$.  Then eq.(\ref{e200}) becomes a second order equation
in $\rho$ where the physical solution easily is extracted.  Now the
effective radial potential depends on both scattering length and
effective range of the two-body interaction.

\paragraph*{Angular eigenvalues.}

\begin{figure}[hbt]
\vspace*{-0mm}
\begin{center}
\psfig{file=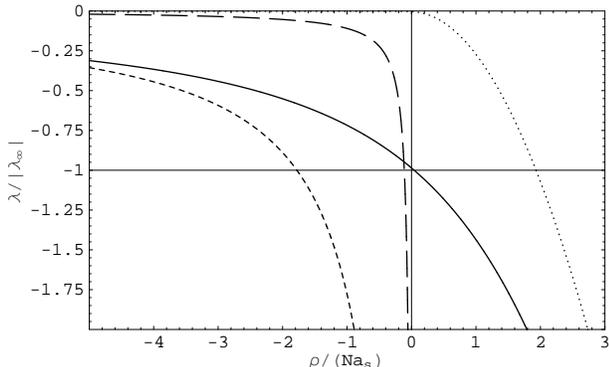,width=8.5cm,%
bbllx=3.0cm,bblly=18.5cm,bburx=15.4cm,bbury=25.1cm}
\end{center}
\vspace*{-0.5cm}
\caption[]{
The angular eigenvalues in units of $\lambda_{\infty} = 
-1.65 N^{7/3}(1-2/N)$ as functions of $\rho/(Na_s)$ obtained from
eq.(\ref{e200}). The computation is for $N=100$. 
The three different terms from eqs.~(\ref{e200}),
i.e. B-term (dotted), $13$-term in eq.(\ref{e90}) (long-dashed) 
$34$-term in eq. (\ref{e100}) (short-dashed), 
are shown individually along with the sum (solid). }
\label{fig1}
\end{figure}

The expression in eq.(\ref{e200}) is straightforward to compute as
a function of $\lambda$ via the (possibly complex) values of $\nu$.  
In fig.~\ref{fig1} we show the
computed function $\rho/a_s$ as a function of $\lambda$ obtained as a
sum of three contributions, i.e. the terms related to 
the Faddeev components
between particles $1-2$ ($B$-term), $1-3$ (eq.~(\ref{e90})) and $3-4$
(eq.~(\ref{e100})).  This procedure is obviously easier than solving
the transcendental equation to get $\lambda$ as a function of $\rho/a_s$.
The immediate implication is that the interaction only enters via the
ratio $\rho/a_s$.

We confine ourselves to the most interesting cases of negative
$\lambda$ corresponding to attractive two-body interactions. The terms
$1-3$ and $3-4$ are both negative and vary from small $\lambda$ and
large $\rho/a_s$ to large $\lambda$ and small $\rho/a_s$. The first
term is much smaller but crosses the $\rho=0$ axis and is therefore
responsible for the same behavior of the sum of the three terms.
However, this very important zero point for $\lambda$ is much larger
for the sum than for the first term.  We used units of
$\lambda_{\infty}$ from \cite{sor04} and a scaling by $N$ on the
$\rho$-axis. Then the figure is fairly independent of particle number.

When both the index and $\lambda$ are small, $\nu \ll 1$, the Jacobi
functions are almost constant allowing the analytic result
\begin{eqnarray}  \label{e210}
  \lambda(\rho)
  \simeq
  \sqrt{\frac2\pi}
  N(N-1)\frac{\gammafkt{3N-3}}{\gammafkt{3N-6}}
  \frac{a_s}{\rho} \;,
\end{eqnarray}
where the validity condition, $\nu\ll1$, can be translated into
$\rho\gg N^{5/2}|a_s|$.  This is the asymptotic behavior seen to the
far left in fig.~\ref{fig1}.  For $N \gg 1$ where our center of mass
separation is less important this result is identical to that derived
in \cite{boh98} for a constant angular wave function and a zero-range
interaction renormalized for use in mean-field computations.

At the other limit of large $\rho/a_s$ and large negative $\lambda$ we
also obtain a closed analytic expression, i.e.
\begin{eqnarray}  \label{e220}
  \lambda(\rho)
  \simeq
  - 2 \frac{\rho^2}{a_s^2} \;,
\end{eqnarray}
where now only the first term in eq.(\ref{e200}) contributes.  In fact
this behavior corresponds precisely to the energy of a two-body bound
state as $E_2 = \hbar^2 \lambda /(2m \rho^2)$.  We emphasize that the
interaction only enters via the ratio $\rho/a_s$.  The scales in
fig.~\ref{fig1} then implies that the results only exhibit the
behavior for distances where $\rho$ is (much) larger than $a_s$.

\begin{figure}[hbt]
\vspace*{-0mm}
\begin{center}
\psfig{file=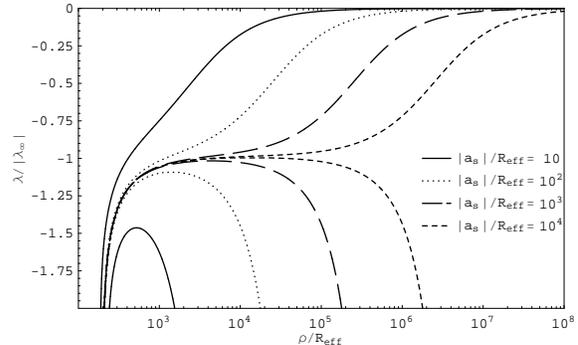,width=8.5cm,%
bbllx=3.0cm,bblly=17cm,bburx=18.8cm,bbury=25.5cm}
\end{center}
\vspace*{-0.5cm}
\caption[]{The angular eigenvalues in  units of $\lambda_{\infty} = 
-1.65 N^{7/3}(1-2/N)$ for $N=100$ as functions of hyperradius 
for several values of the scattering length $a_s$ both in units of the
effective range $R_{eff}$. }
\label{fig2}
\end{figure}

For many realistic systems the interesting region is $\rho/|a_s| < 1$,
i.e. the region on fig.~\ref{fig1} where $\rho \approx 0$.  
To exhibit the behavior in this region we need to include the effective
range in eq.(\ref{e140}) and solve the resulting second order equation
in $\rho/R_{eff}$.  The resulting angular eigenvalues for different
scattering lengths are shown in fig.~\ref{fig2} as a function of
hyperradius. No real solutions exist at small values of $\rho$ where
the detailed behavior of the two-body interaction in any case is
important. Higher order terms in the effective range expansion would
also allow the small $\rho$-values.

As $\rho$ increases the eigenvalue levels off when $a_s$ is
sufficiently large.  The height of this plateau is independent of
$a_s$ and in fact precisely equal to the value of $\lambda$ obtained
from fig.~\ref{fig1} for $\rho=0$. This value is 
rather accurately given by the unit used in fig.~\ref{fig1},
$\lambda_{\infty} = -1.65 N^{7/3}(1-2/N)$, extracted by elaborate
numerical solutions of a variational equation with finite-range
interactions \cite{sor03}.

As $\rho$ increases further to values exceeding $a_s$ the eigenvalues
either bend up and approach zero ($a_s<0$) as expressed in eq.(\ref{e210}) 
or bend down diverging ($a_s>0$) as expressed in eq.(\ref{e220}). These two
characteristics then reflect the different signs of the scattering
length and the related ability of the two-body potential to support a
bound state or not.

\begin{figure}[hbt]
\vspace*{-0mm}
\begin{center}
\psfig{file=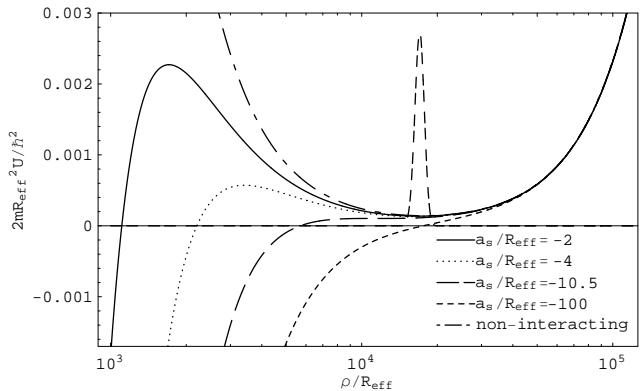,width=7cm,%
bbllx=5cm,bblly=17.5cm,bburx=16cm,bbury=25cm}
\end{center}
\vspace*{-0.4cm}
\caption[]{The effective radial potential in eq.(\ref{e120}) for $N=100$ 
as a function of $\rho$ for several values of the scattering length
$a_s$ both in units of the effective range $R_{eff}$. 
The oscillator length is $b_t=1442R_{eff}$ as in \cite{sor02a}. 
The radial wave
function (two-dashed line) located in the minimum is shown for $a_s = -2
R_{eff}$. }
\label{fig3}
\end{figure}

\paragraph*{Radial potential and solutions.}

The angular eigenvalues are now inserted into the effective radial
potential in eq.(\ref{e120}). The behavior for different scattering
lengths is seen in fig.~\ref{fig3}. The confining trap provides the
large positive potential for large distances.  The minimum appearing
for relatively weakly attractive potentials of $|a_s|/R_{eff}\lesssim 10.5$
gradually disappears as the attraction increases. The barrier towards
small distance disappears roughly when $|a_s| N /b_{t}> 0.67$ as
derived previously in \cite{boh98,sor03}. The improvement is towards
less stability decreasing with increasing $N$.  The minimum supports
quasistationary states with characteristic features similar to
condensates. The radial wave function $f$ obtained from
eq.(\ref{e110}) is shown in fig.~\ref{fig3} as a distribution
around the minimum. Other interaction parameters leaving the minimum
would produce almost indistinguishable radial wave functions.

The plateau region for large $a_s$ in fig.~\ref{fig2} now appears in
fig.~\ref{fig3} as a $\rho^{-2}$-potential with a strength of
approximately $2.25 N^2 - 1.65 N^{7/3}$, which is negative already
when $N>3$.  As this strength is less than $-0.25$ the Efimov
conditions are fulfilled and a number of states related by simple
scaling properties are solutions to the radial equation in
eq.(\ref{e110}), see \cite{nie01,sor02}. These states are located in
the plateau region far outside the range of the two-body interaction
but before the confining wall of the trap.  If created they may have a
sufficiently long lifetime to be seen or perhaps play a role in some
processes.  They have obviously enough energy to decay into bound
cluster states of much smaller size.

\paragraph*{Conclusions.}

Zero-range interactions have been used extensively for two and
three-body systems. It is a substantial simplification and very
accurate for large distances compared to the range of the potential.
We extend the application to $N$-boson systems by use of the adiabatic
hyperspherical expansion method. We expect that two-body correlations
are most important and dominated by $s$-waves. As zero-range
interactions only are active in $s$-waves we use accordingly wave
functions consisting of only $s$-wave Faddeev two-body
amplitudes.  This may then be viewed as the largest contribution in an
expansion in both partial waves and many-body correlation 
amplitudes \cite{sor04}.
For dilute systems we thereby obtain an accurate effective radial
potential from a transcendental algebraic equation.

We derive the crucial equation and renormalize the zero-range
interaction by an effective range expansion in terms of two-body
phase-shifts. This is analogous to the field theoretical
renormalization studied intensively for three-body systems.  We
extract and discuss the pertinent general scaling properties for both
attractive and repulsive interactions.  Use of parameters
corresponding to systems forming condensates reveal the established
properties and stability conditions.

The method has many applications and extensions which are beyond the
scope of this letter, e.g. more systematic computations of various
properties like one- and two-body densities, studies of correlations
or effects beyond the mean-field approximation, investigations of
dynamics in general and in particular the recombination process into
bound cluster states, applications on more complicated systems like
two-component boson systems, extensions to fermionic and mixed systems.

\end{document}